\documentclass[aps,prl,twocolumn,showpacs,groupedaddress]{revtex4}

\usepackage{bm}
\usepackage{graphicx}

\newcommand{\etal}{{\it et al.}}%
\newcommand{\SRO}{{Sr$_{2}$RuO$_{4}$}}%
\newcommand{\CRO}{{Ca$_{2}$RuO$_{4}$}}%

\begin{document}


\title{
From Mott insulator to ferromagnetic metal: a pressure study of 
Ca$_{\bm{2}}$RuO$_{\bm{4}}$
}

\author{Fumihiko Nakamura}
\affiliation{
Department of Quantum Matter, ADSM, Hiroshima University,
Higashi-Hiroshima 739-8530, Japan}
\author{Tatsuo Goko}
\affiliation{
Department of Quantum Matter, ADSM, Hiroshima University,
Higashi-Hiroshima 739-8530, Japan}
\author{Masakazu Ito}
\affiliation{
Department of Quantum Matter, ADSM, Hiroshima University,
Higashi-Hiroshima 739-8530, Japan}
\author{Toshizo Fujita}
\affiliation{
Department of Quantum Matter, ADSM, Hiroshima University,
Higashi-Hiroshima 739-8530, Japan}
\author{Satoru Nakatsuji}
\affiliation{
Department of Physics, Kyoto University, 
Kyoto 606-8502, Japan}
\author{Hideto Fukazawa}
\affiliation{
Department of Physics, Kyoto University, 
Kyoto 606-8502, Japan}
\author{Yoshiteru Maeno}
\affiliation{
Department of Physics, Kyoto University, 
Kyoto 606-8502, Japan\\
Kyoto University International Innovation Center, 
Kyoto 606-8501, Japan\\
CREST, Japan Science and Technology Corporation, Japan}
\author{Patricia Alireza}
\affiliation{
Cavendish Laboratory University of Cambridge,
Madingley Road CB3 OHE, Cambridge, United Kingdom}
\author{Dominic Forsythe}
\affiliation{
Cavendish Laboratory University of Cambridge,
Madingley Road CB3 OHE, Cambridge, United Kingdom}
\author{Stephen R. Julian}
\affiliation{
Cavendish Laboratory University of Cambridge,
Madingley Road CB3 OHE, Cambridge, United Kingdom}

\date{\today}


\begin{abstract}
We show that the pressure-temperature phase diagram of the Mott insulator 
\CRO{} features a metal-insulator 
transition at 0.5GPa: at 300K from paramagnetic insulator 
to paramagnetic quasi-two-dimensional metal; at $T \leq$ 12K from 
antiferromagnetic insulator to ferromagnetic, highly anisotropic,  
three-dimensional metal.
We compare the metallic state to that of the structurally related $p$-wave 
superconductor \SRO{}, and discuss the importance of 
structural distortions, which are expected to couple strongly to pressure.
\end{abstract}


\maketitle


The Mott transition from the metallic to the insulating state, driven by a coulomb interaction, is one of the most dramatic and fundamental many-body effects in condensed matter physics.
In recent years there has been a resurgence of interest in the Mott transition due on the one hand to the discovery of materials such as the high-$T_c$ cuprates and the manganites in which the metal-insulator transition plays an essential role \cite{imada}, and on the other to new insights gained from theoretical developments such as the dynamical mean field theory \cite{georges} and the recognition of the importance of orbital ordering at the metal-insulator transition \cite{imada}. 

The system Ca$_{2-x}$Sr$_{x}$RuO$_4$ has recently been investigated in detail and shown to have a metal-insulator transition at 0K at $x\sim 0.2$ \cite{SN_CSRO}.
This is a layered perovskite system, isostructural to the cuprate La$_{2-x}$Sr$_{x}$CuO$_4$, with one end-member, Sr$_2$RuO$_4$, being a highly anisotropic metal with an unconventional (probably $p$-wave) superconducting ground state \cite{SRO_YM,maeno_pt}, 
while the other, Ca$_2$RuO$_4$,  is an antiferromagnetic (AF) insulator for $T < 113$K, a paramagnetic (PM) insulator for $113$K$<T<360$K, and a \lq\lq bad metal" for $T>360$K \cite{SN_CSRO,cao_1}. 

This system is of great interest for the following reasons:

Firstly, the overwhelming majority of work on the metal-insulator transition has been done on $3d$-transition metal compounds (mainly oxides) such as V$_2$O$_3$ \cite{imada}.  
$3d$-orbitals are comparatively compact and close to the ionic core, giving rise naturally to small bandwidths $W$, and large Coulomb repulsion $U$ between $3d$-electrons on the same site.  
In contrast, $4d$ metal-insulator systems are rare, but they should offer a different viewpoint because the more extended $4d$-orbitals will tend to have larger effective correlation energy $U/W$, but greater susceptibility to orbital ordering via structural distortions around the $4d$-site.  
Indeed the phase diagram of Ca$_{2-x}$Sr$_x$RuO$_4$ shows a series of structural transitions that are intimately related to changes in the electronic state \cite{SN_CRO}.

Secondly, although the cuprates show superconductivity in close proximity to a Mott insulating state, a clear connection between the metal-insulator transition and superconductivity has not been made experimentally, and there are many unresolved issues associated with the physics of the cuprates near the Mott insulating state.  As a quasi-two-dimensional (Q2D) system bridging the gap between a Mott insulator and an unconventional superconductor, the Ca$_{2-x}$Sr$_{x}$RuO$_{4}$ system promises to shed light on this question. 

Finally, detailed understanding of how $p$-wave superconductivity arises in Sr$_2$RuO$_4$ is lacking.  $p$-wave superconductivity is assumed to be mediated by magnetic fluctuations at low $q$, normally associated with proximity to a ferromagnetic (FM) state at 0K (i.e.\ to a FM quantum critical point), but Sr$_2$RuO$_4$ does not itself appear to be close to FM order. 
The doped system, however, at $x \sim 0.5$ is on the verge of FM order at 0K \cite{SN_CRO}, although long-range FM order does not develop for any value of $x$. 

Pressure is generally to be preferred to chemical doping to tune the properties of materials because pressure does not introduce disorder. 
Disorder can be extremely hostile to unconventional superconducting states \cite{mackenzie_prl} 
and it also introduces enormous complications in studying the Mott transition. 
In this paper we show that application of very modest pressures to pure Ca$_2$RuO$_4$ transforms it from an insulator to a highly anisotropic metal with a FM metallic ground state, and we follow the evolution of this ground state as a function of pressure. 
Our results shed new light on the nature of ferromagnetism in the layered ruthenates, and open up a new $4d$-system for the study of the metal-insulator transition in the absence of disorder.


We used single crystal \CRO\ with an essentially stoichiometric oxygen content (S-CRO) \cite{SN_CRO}.
Resistivity under pressure $P$ was measured by a standard four-probe method, with a ring contact geometry for the $c$-axis resistivity $\rho_{c}$. 
The electrodes were made by a silver evaporation technique. 
The typical sample size was $0.15\times 0.5\times 0.04$ mm$^3$ for in-plane resistivity $\rho_{ab}$, and $0.35\times 0 .35\times 0.04$ mm$^3$ for $\rho_{c}$. 
Pressures up to 8GPa were generated in a cubic anvil device with an equal volume mixture of Fluorinert FC70 and FC77 (3M Co. St. Paul, MN, USA) as a pressure-transmitting medium. 
In this device quasi-hydrostaticity can be generated by isotropic movement of six anvil tops even after the fluid medium vitrifies at low temperature and high pressure.

Figure 1 shows $\rho_{ab}$ and $\rho_c$ at 300K as a function of pressure up to 8 GPa. 
Both $\rho_{ab}$ and $\rho_{c}$ drop discontinuously at 0.5 GPa, indicating an insulator-metal transition. 
The discontinuity indicates that the transition is of first order and is thus most likely accompanied by a structural change \cite{friedt_unpub}.

\begin{figure}
\begin{center}
\includegraphics[width=65mm,clip]{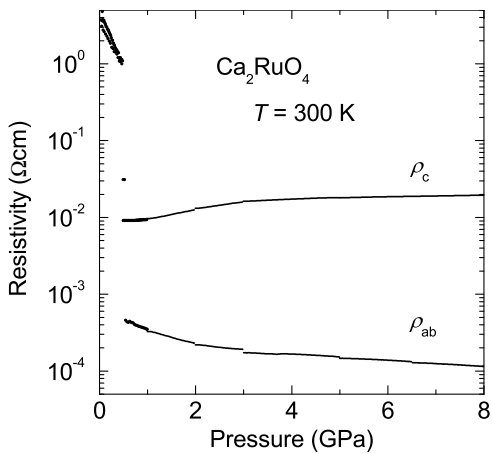}
\caption{
Pressure variation of $\rho_{ab}$ and $\rho_{c}$ at 300 K. 
The insulator-metal transition occurs at 0.5 GPa, 
where  $\rho_{ab}$ and $\rho_{c}$ fall 
by approximately four and two orders of magnitude, respectively.
}
\label{Fig1.eps}
\end{center}
\vspace{-0.5cm}
\end{figure}

In the insulating state both $\rho_{ab}$ and $\rho_{c}$ decrease isotropically and monotonically with increasing pressure. 
Following the discontinuous drop at 0.5 GPa however, while $\rho_{ab}$ continues to fall, $\rho_{c}$ begins to {\em increase}.  
The anisotropy ratio, $\rho_{c}$/$\rho_{ab} \sim 2$ in the insulating phase, jumps to about 100 in the metallic phase. 
Thus, the Q2D nature of the metallic state is comparable with high $T_c$ systems or \SRO. 

Figure 2 shows $\rho_{ab}$ and $\rho_{c}$ as a function of temperature at several fixed pressures. 
At ambient pressure, both $\rho_{ab}$ and $\rho_{c}$ show an activation-type increase.
In contrast, above 0.5 GPa \CRO\ shows a Q2D-metallic nature for temperatures above $\sim$25K. 
That is, $\rho_{ab}(T)$ shows metallic (d$\rho_{ab}$/d$T>$0) conduction whereas $\rho_{c}$ has a negative slope indicating nonmetallic conduction. 
The anisotropy ratio rises to $\sim$1000 at 25K. 
The nonmetallic behaviour of $\rho_{c}$ down to 25K in pressurised \CRO\ is in sharp contrast with the metallic $\rho_{c}$ seen below 130K in \SRO. 
The rise in $\rho_c$ with increasing pressure at 300K also contrasts with simple ideas of metallic conduction, in which the bandwidth would increase with pressure; but the same effect is seen in \SRO\ at $T>130$K \cite{yoshida}, so this seems to be a signature of incoherent $c$-axis transport in the ruthenates. 
We infer that above 25K the $c$-axis conduction in metallic \CRO is mainly described by a tunnelling mechanism as in the high $T_c$'s.

\begin{figure}
\begin{center}
\includegraphics[width=70mm,clip]{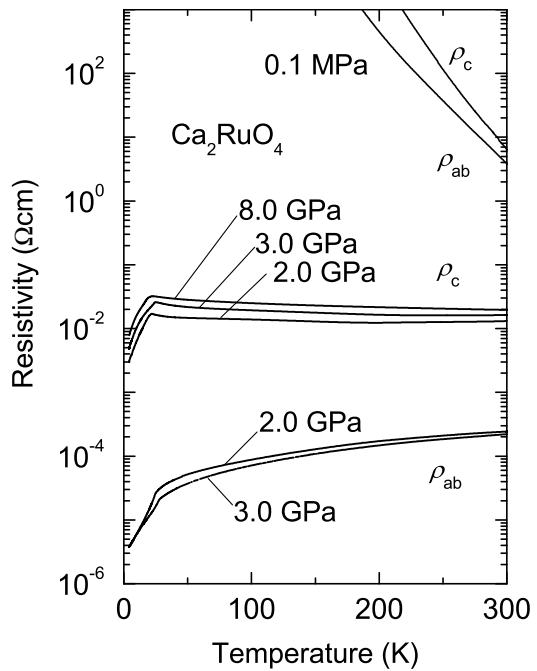}
\caption{
Temperature variation of $\rho_{ab}$ and $\rho_{c}$ 
at fixed pressure $P$. 
For $P> 0.5$ GPa, $\rho_{ab}(T)$ is metallic 
while $\rho_{c}(T)$ above $T^* \sim 25$K exhibits 
nonmetallic temperature dependence. 
}
\label{Fig2.eps }
\end{center}
\vspace{-0.5cm}
\end{figure}

As the temperature falls below $\sim$25K $\rho_{c}(T)$ peaks and then decreases rapidly, indicating a change from two-dimensional to three-dimensional metallic conduction. 
At the temperature $T^*$ corresponding to the kink in the $\rho_{c}(T)$ curves, $\rho_{ab}(T)$ also exhibits an abrupt change in slope, falling rapidly to a residual resistivity of $\sim 3 \mu \Omega$ cm. 
This is a very low resistivity for an oxide metal, indicating that our samples are of high quality, but it is still somewhat higher than is required to observe superconductivity in \SRO\ \cite{mackenzie_prl}.
The behaviour near 25K is reminiscent of a phase transition. 
In order to clarify this point, the field ($H$) and temperature ($T$) dependence of the magnetization have thus been measured under pressures of 0.1 MPa, 0.2, 0.5, 0.7, and 0.8 GPa using a piston-cylinder Be-Cu clamp cell with a commercial SQUID magnetometer (Quantum Design, model MPMS) \cite{HPcell}. 
We have used several rectangular pieces of sample with total mass 12.5mg for $H_{\perp c}$ and 6mg for $H_{\| c}$.

Figure 3 shows magnetization ($M$) curves at 2 K with $H_{\perp c}$. 
FM 
ordering is indicated by the hysteresis in the $M$-$H$ curves at 0.7 GPa, a pressure at which the system is metallic in the corresponding resistivity measurements. 
In contrast, no hysteresis is observed in the $M$-$H$ curves at 0.1 MPa and 0.2 GPa where \CRO\ is in the AF insulating phase. 
The remnant magnetisation is 0.09 $\mu_{\rm B}$ at 0.7 GPa, growing to 0.125 $\mu_{\rm B}$ at 0.8 GPa \cite{note2}, 
while the coercive force of $H_{\rm c} \sim$ 50mT, which is defined as the field necessary to restore zero magnetisation, increases only weakly.

\begin{figure}
\begin{center}
\includegraphics[width=65mm,clip]{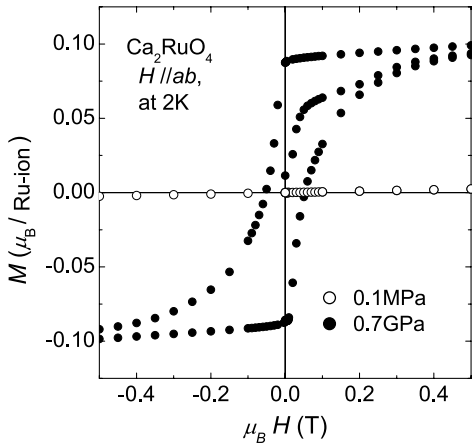}
\caption{
Magnetisation curves for $P=$ 0.7 GPa and 0.1 MPa, measured at 2 K. 
Hysteresis indicating FM ordering is observed at 0.7 GPa, 
but not at 0.1 MPa. 
The remnant $M_{r}$ is 0.09 $\mu_{\rm B}$ at 0.7 GPa 
while the coercive force is $\mu_{\rm B}H_{\rm c} \sim 50$ mT.
}
\label{Fig3.eps}
\end{center}
\vspace{-0.5cm}
\end{figure}

The volume fraction of the AF insulating phase ought to decrease with pressure above 0.5 GPa, 
so the remnant magnetisation originates from the metallic phase. 
Moreover the small moment compared with the saturated moment 2$\mu_{\rm B}$ of a localized Ru$^{4+}$ ion implies itinerant magnetism. 
The shape of the hysteresis curve, which indicates a typical soft ferromagnet, is different from that for the canted AF case seen for example in the AF insulating phase of \CRO\ with excess oxygen (L-CRO) \cite{SN_CRO}. 
We have also observed a similar shape of hysteresis loop with $H_{\| c}$ at 0.8 GPa where the remnant magnetisation is 0.08 $\mu_{\rm B}$ and the coercive force of $H_{\rm c}$ is $\sim 50$mT. 
Comparing the hysteresis measured in $H_{\| c}$, we do not find anisotropy typical of antiferromagnetism. 
We therefore deduce that \CRO\ in the pressure-induced metallic phase is an itinerant ferromagnet at low $T$, although neutron scattering should be done to confirm this. 

Figure 4 summarises the pressure dependence of $T_{\rm N}$ and $T_{\rm C}$ obtained from our $M(T)$ curves, and of $T^{\rm *}$ obtained from resistivity. 
$T_{\rm N}$ is almost independent of pressure up to 0.5GPa, 
where it abruptly vanishes and is replaced by FM order below $\sim$12K. 
The FM $T_{\rm C}$ then rises with increasing pressure, becoming comparable to $T^*$ near 1GPa. 
Thus, the remarkable drop in $\rho(T)$ at $T^*$ is most probably due to reduced magnetic scattering at a FM ordering transition. 
$T^*$ continues to rise, reaching 25K at $\sim$5 GPa, then it decreases gradually. 
Extrapolation of the $T^*$ vs $P$ curve suggests that the FM order may be completely suppressed at $\sim$15GPa.

\begin{figure}
\begin{center}
\includegraphics[width=65mm,clip]{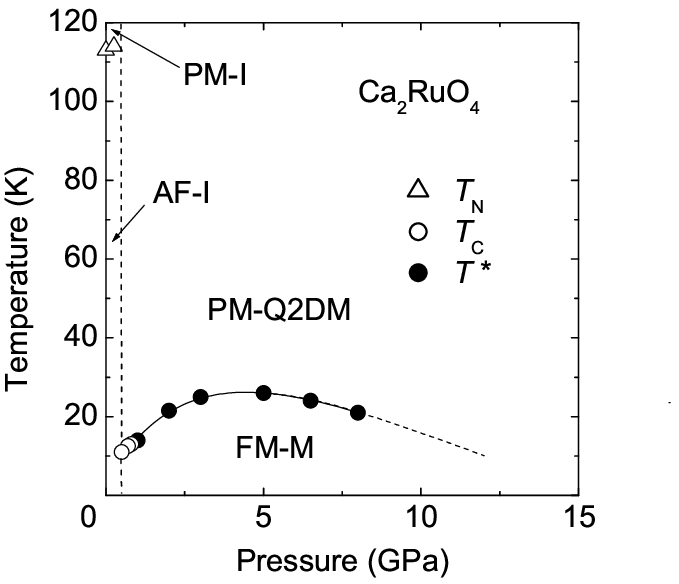}
\caption{
The characteristic temperatures $T_{\rm N}$ (AF ordering), 
$T_{\rm C}$ (FM ordering) and $T^*$ (cusp in the resistivity) 
plotted against pressure.  
The phases are labelled by:  
PM-I = paramagnetic insulator, AF-I = antiferromagnetic insulator, 
PM-Q2DM = paramagnetic quasi-two-dimensional metal, 
FM-M = ferromagnetic metal. 
Solid and dotted lines are guides to the eye. }
\label{Fig4.eps}
\end{center}
\vspace{-0.5cm}
\end{figure}

From $\sim$360 to 113K the 0.5 GPa transition is from a PM insulator to a PM-Q2D metal. 
The PM insulating phase with thermally activated conductivity is important because it shows that \CRO\ is a many-body insulator, as opposed to being a band-insulator with fully opened gaps on the Fermi surface due to AF order. 
It is also instructive to contrast the PM metallic state with the metallic phase of the manganites, which are always FM and can only be reached by doping.
This reflects the difference in on-site repulsion $U$ in $3d$ and $4d$ systems.
Mn has a large $U$, so the Mn ion in the manganites is in a high-spin configuration, and the strong Hund's rule coupling requires that the spin of a conduction electron in an $e_g$ orbital must be parallel to the localised $t_{2g}$ spins both on the ion it is hopping from and the one it is hopping to (so-called \lq\lq double exchange"). 
In the ruthenates in contrast lower $U$ produces a low-spin configuration, and there is no constraint on the spin-direction of the hopping electron in the metallic state. 

It is tempting to interpret the phase diagram of \CRO\ as a classic Mott transition to a metal with a FM ground state \cite{Fresard}. 
But this might be too simple. 
Analogy with Ca$_{2-x}$Sr$_x$RuO$_4$ shows that the physics is not simply driven by changes in $U/W$, but also the orbital degeneracy of the Ru$^{4+}$ $t_{2g}$ levels may abruptly change due to structural transitions.
These have been well characterised by neutron and x-ray scattering \cite{SN_CSRO,MB_CSRO}.
Pure \SRO\ is tetragonal, but at the Ca rich end the structure is distorted by rotations of the oxygen octahedra about the $c$ axis, tilts about an in-plane axis, and large variations in the degree of \lq flattening' of the octahedra along the $c$ axis. 
In pure \CRO\ the $T>360$K metallic phase is L-Pbca, with a smaller volume and weaker tilt and flattening distortions than the $T<360$K insulating S-Pbca phase. 
Pressure should stabilise the low volume phase, so we expect the structural transition associated with the first-order insulator-metal transition to be from L-Pbca to S-Pbca \cite{friedt_unpub}. 
Fang and Terakura \cite{Fang} have performed electronic structure calculations that take account of the distortions. 
They find that flattening, which lowers the $d_{xy}$ orbital relative to the $d_{xz}$ and $d_{yz}$ orbitals, favours an insulating AF ground state; the tilt stabilises AF correlations in the metallic state; while rotations favour ferromagnetism. 
Thus the insulator-metal transition and AF order may couple separately to pressure, the former caused by suppression of flatting, the latter by reduced tilting. 

We now turn to a discussion of the FM metallic state above the insulator-metal transition. 
According to band structure calculations FM order in layered ruthenates is a reflection of rotational distortions \cite{Fang,Hotta}, 
which develop FM coupling by enlarging the density of states at the Fermi energy via band narrowing. 
This idea is supported by the large FM correlations in Ca$_{2-x}$Sr$_x$RuO$_4$ with $x =$ 0.5, in which the octahedra are rotated but not tilted \cite{unpub}. 
More generally \cite{mazin}, ruthenates show a tendency to ferromagnetism due to an unusually large admixture at the Fermi surface of the oxygen $p$-orbitals, which can gain exchange energy from FM order. 
It will be very interesting to have electronic structure calculations on metallic \CRO, once the crystal structure is known accurately in the high-pressure phase. 
It will also be interesting to compare the properties of the itinerant-FM Q2D-metal with theoretical predictions \cite{Hatatani}. 

Finally, the large range of temperatures for which pressurised \CRO\ exhibits a non-metallic $\rho_c$ is of interest. 
In Ca$_{2-x}$Sr$_{x}$RuO$_4$, for 0.5 $<x<$ 1.5 where the octahedra are rotated but not tilted, nonmetallic $\rho_{c}(T)$ is also seen  over a wide temperature range,  while \SRO, having non-rotated octahedra, exhibits metallic $\rho_{c}$ below 130K. 
We therefore suspect that the rotational distortion affects not only the magnetic properties but also the two-dimensionality in the electronic transport in the metallic phase.

In conclusion, high-purity \CRO\ is a $4d$-system that at 0.5 GPa has a transition from a Mott insulating state to a metal with a FM ground state.  
The $P$-$T$ phase diagram of this system is unique and quite different from that of the doped system \cite{SN_CSRO,HF_LCRO}.
Furthermore, our results suggest the existence of a FM quantum critical point at pressures above 10 GPa. 
We thus have strong interest in measurements at higher pressures as a crucial test of the connection between the $p$-wave superconductivity of \SRO, and the ferromagnetism of metallic \CRO.

We thank T.Moriya, G.G. Lonzarich, A.P. Mackenzie and D.J. Singh for informative discussions. 
A part of this work was supported by a Grant-in-Aid for Scientific Research 
on Priority Areas, from the MECSST of Japan, and by the EPSRC of Great Britain. 
%


\bibliography{basename of .bib file}

\begin{thebibliography}{99}
%
\bibitem{imada}
for a review see M. Imada, A. Fujimori and Y. Tokura, 
Rev. Mod. Phys. {\bf 70}, 1039 (1998).
%
\bibitem{georges}
see e.g. A. Georges, G. Kotliar, W. Krauth and M. J. Rozenberg, 
Rev. Mod. Phys. {\bf 68}, 13 (1996).
%
\bibitem{SN_CSRO}
S. Nakatsuji and Y. Maeno, 
Phys. Rev. Lett. {\bf 84}, 2666 (2000); 
Phys. Rev. B {\bf 62}, 6458 (2000).
%
\bibitem{SRO_YM}
Y. Maeno \etal, 
Nature {\bf 372}, 532 (1994).
%
\bibitem{maeno_pt}
Y. Maeno, T. M.  Rice, M. Sigrist,
Physics Today {\bf 54}, 42 (2001).
%
\bibitem{cao_1}
G. Cao \etal, 
Phys. Rev. B {\bf 56} (1997) R2916.
%
\bibitem{SN_CRO}
S. Nakatsuji \etal, 
J. Phys. Soc. Jpn. {\bf 66}, 1868 (1997).
%
\bibitem{mackenzie_prl}
A. P. Mackenzie \etal, 
Phys. Rev. Lett. {\bf 80},161 (1998). 
%
\bibitem{friedt_unpub}
This has very recently been confirmed by neutron and x-ray 
diffraction measurements; 
O. Friedt et al., to be published. 
%
\bibitem{yoshida}
K. Yoshida \etal, Phys. Rev. B {\bf 58}, 15062 (1998). 
%
\bibitem{HPcell}
Y. Uwatoko \etal, Rev. High Press. Sci. Tech.,{\bf 7}, 1508 (1998).  
%
\bibitem{note2}
We note here the intrinsic value of the remnant magnetisation could be 
larger than the observed ones because pressure of 1GPa might be too small 
to eliminate the AF insulator domain completely. 
%
\bibitem{Fresard}
R. Fresard and G. Kotliar, 
Phys. Rev. B {\bf 56}, 12909 (1997).
%
\bibitem{MB_CRO}
M. Braden \etal, 
Phys. Rev. B {\bf 58}, 847 (1998).
%
\bibitem{MB_CSRO}
O. Friedt \etal, 
Phys. Rev. B {\bf 63}, 174432 (2001).
%
\bibitem{Fang}
Z. Fang and K. Terakura, 
Phys. Rev. B {\bf 64}, 020509(R) (2001).
%
\bibitem{Hotta}
Another possible reason for the occurrence of FM order is discussed based on the orbital degree of freedom (see T.Hotta and E.Dagotto, cond/mat0108531).
%
\bibitem{unpub}
S. Nakatsuji \etal, 
(unpublished).
%
\bibitem{Tokura}
R. Matzdorf \etal, 
Science {\bf 289}, 746 (2000).
%
\bibitem{HF_LCRO}
H. Fukazawa and Y. Maeno, 
J. Phys. Soc. Jpn. {\bf 70}, 460 (2001).
%
\bibitem{mazin}
I. I. Mazin and D. J. Singh, 
Phys. Rev. Lett. {\bf 82}, 4324 (1999)
%
\bibitem{Hatatani}
M. Hatatani \etal, 
J. Phys. Soc. Jpn. {\bf 64}, 3434 (1995). 
%
\end{thebibliography}


\end{document}